\documentclass[preprint2]{aastex701}
\begin{document}

\title{Benchmarking Astrochemistry Paradigms: Relative Absence of $\rm{C_6H_5CN^+}$ in the Diffuse ISM}

\author[orcid=0000-0001-8803-3840]{Daniel Majaess}
\affiliation{Department of Chemistry and Physics, Mount Saint Vincent University, Halifax, Nova Scotia, B3M2J6 Canada.}
\email[show]{Daniel.Majaess@msvu.ca}

\author[orcid=0000-0002-8746-9076]{Cercis Morera-Boado}
\affiliation{IXM-Secihti-Centro de Investigaciones Químicas, IICBA, Universidad Autónoma del Estado de Morelos, Cuernavaca, 62209, Morelos, Mexico.}
\email[]{cermor@gmail.com}

\author[orcid=0000-0003-3469-8980]{Tina A. Harriott}
\affiliation{Department of Mathematics and Statistics, Mount Saint Vincent University, Halifax, Nova Scotia, B3M2J6 Canada.}
\email[]{Tina.Harriott@msvu.ca}

\author[]{Quazi Rahi}
\affiliation{Department of Mathematics and Statistics, Mount Saint Vincent University, Halifax, Nova Scotia, B3M2J6 Canada.}
\email[]{Quazi.Rahi@msvu.ca}

\author[]{Halis Seuret}
\affiliation{Prepa TEC Campus Cuernavaca, Instituto Tecnológico y de Estudios Superiores de Monterrey, 62209, Morelos, Mexico.}
\email[]{halisseureth@gmail.com}

\author[orcid=0000-0001-6662-3428]{Lou Massa}
\affiliation{Hunter College \& the PhD Program of the Graduate Center, City University of New York, New York, USA.}
\email[]{lmassa@hunter.cuny.edu}

\author[orcid=0000-0001-8397-5353]{Ch\'erif F. Matta}
\affiliation{Department of Chemistry and Physics, Mount Saint Vincent University, Halifax, Nova Scotia, B3M2J6 Canada.}
\affiliation{Department of Chemistry, Saint Mary's University, Halifax, Nova Scotia, B3H3C3 Canada.}
\affiliation{D\'epartement de Chimie, Universit\'e Laval, Qu\'ebec, G1V0A6 Canada.}
\affiliation{Department of Chemistry, Dalhousie University, Halifax, Nova Scotia, B3H4J3 Canada.}
\email{[]Cherif.Matta@msvu.ca}

\begin{abstract}
The detectability of $\rm{C_6H_5CN^+}$ (benzonitrile cation) in the diffuse ISM is re-evaluated. A holistic evidentiary framework suggests $\rm{C_6H_5CN^+}$ is relatively absent in the diffuse ISM owing to the following concurrently: a marginal intramolecular vibrational energy redistribution (IVR) favoring fragmentation, recurrent fluorescence being an improbable mechanism in this case to prevent dissociation, unceasing photon strikes, mismatches between observed DIBs and experimental results, and the hitherto absence of DIBs matching any similarly sized cations. The putative gap in bottom-up synthesis is reaffirmed (diffuse ISM), and although DIB sources are largely unknown, within a broader approach the lines can help benchmark astrochemistry paradigms.  The results relied on new advantageous \citeauthor{da24} experimental spectra, an expanded observational DIB analysis (APO catalog), and complementary $\omega$B97X-D/cc-pVTZ computations.
\end{abstract}

\keywords{\uat{Astrochemistry}{75}}

\section{Introduction}
Diffuse interstellar bands (DIBs) are a long-standing spectroscopic mystery \citep{heg22}, and are tied to unknown molecules superposing absorption lines along the sightline.  Less than 1\% of $>550$ DIBs are associated with a source, and in that case the sole carrier is contested \citep[C$_{60}^{+}$, see \S \ref{sec:conclusion}, and e.g.,][]{ca15,gal21,ma25a,maj35c}. Concurrently, experimental spectra, quantum chemistry predictions, and DIB catalogs can help benchmark proposed carriers. Toward that end \citet{da24} provided optimal laboratory spectra for the benzonitrile cation ($\rm{C_6H_5CN^+}$, also referred to as BZN$^+$).  The neutral variant was detected in cold dense molecular clouds \citep{mcg18,cer23}, which inevitably motivated a comparison between the cation and DIBs. Cations are probable in the diffuse ISM owing to a low-density and unshielded environment, where higher energy photons are abundant and comparatively unobscured, thus leading to ionization \citep[e.g.,][]{cr85}. Moreover, cations can produce lines in the optical where most DIBs reside, whereas certain neutral counterparts typically absorb shorter wavelengths \citep[e.g.,][]{wei03,ha25,se25}.

Smaller aromatics are particularly susceptible to dissociating higher energy photons present in the diffuse ISM (e.g., $6-13.6$ eV, the interstellar radiation field, ISRF). Consequently, bottom-up synthesis for benzene ring(s) from smaller fragments may be unlikely therein \citep[e.g.,][]{al96}, owing to the environment's comparative transparency, and since ion-neutral induced dipole interactions govern simpler assembly (e.g., CH$\textsuperscript{+}$, CH, CN, etc.).  Benzonitrile features a low intramolecular vibrational energy redistribution (IVR) \citep[e.g., Fig.~2 in][and Fig.~3 in \citealt{mo13}, see also Rice-Ramsperger-Kassel-Marcus theory, RRKM, \citealt{mar52,dig15}]{al96}, and the balance between destruction and creation is biased toward the former.  Limited vibrational density of states inherent to small molecules yields slower IVR, thus energy funnels more quickly through dissociation. Specifically, \citet{al96} concluded that aromatics containing less than 50 carbon atoms will be comparatively absent in the diffuse ISM, and \citet{mo13} support that assertion noting such molecules shall be fully dehydrogenated (e.g., benzonitrile, which has merely 7 carbon atoms). However, and importantly, it was posited in the literature that $\rm{C_6H_5CN^+}$ could survive that destruction via recurrent fluorescence, whereas a separate view is expressed here (\S \ref{sec-RF}). Nevertheless, IVR rapidly occurs for larger PAHs or fullerenes, as energy is redistributed amongst numerous modes before fragmentation occurs, hence the relative stability of C$_{60}^{+}$ against ISRF.  Pertinently, the absence of a detectable bottom-up benzene ring(s) generation scenario in the diffuse ISM is tentatively supported by C$_{60}^{+}$ being the sole carrier for a few DIBs (see \S \ref{sec:conclusion}), and a hitherto lack of convincing matches between DIBs and smaller or intermediate-sized PAH cations \citep[e.g.,][]{sa99}. For sizable PAHs detectability prospects are counterbalanced by increasing isomer populations \citep[e.g.,][]{dia82}, thus tentatively hinting why identifications have yet to emerge.  Alternatively, \citet{kw22} and \citet{ks23} highlight pitfalls endemic to the canonical PAH hypothesis, and champion MAONs (mixed aromatic/aliphatic organic nanoparticles).

The C$\equiv$N radical is of particular interest granted \citet{ma25b} identified its vibrational signature in a histogram of DIB energy differences, and which remained prominent post noise analysis (500-4000 cm$^{-1}$ range).  DIBs may represent a vibrational progression, where energy differences between interrelated DIBs sharing a common carrier may reveal infrared vibrational transitions \citep[e.g.,][]{je93,bon20}.  \citet{ma25b} proposed that a histogram tied to energy differences between highly correlated DIB pairs across the entire APO catalog may broadly reveal bonds endemic to the sources (e.g., aromatics, oop C$-$H bending, C$-$H stretch, C$\mathbf{^{\underline{...}}}$C in-ring, overtones and combinations), thus circumventing challenging comparisons to individual molecules. For example, the \citet{ma25b} analysis exposed the signatures of aromatics and C$\equiv$N (potentially nitriles given the former).  \citet{ma25b} conveyed the effort was a first macro exploratory step requiring independent validation and continuous fine-tuning. The analysis relied on the Pearson$-r$ correlation, which characterizes the linear dependence between separate DIBs, whereby $-1$ is a perfect inverse correlation, zero is no correlation, and $+1$ is an ideal positive correlation ($r = \sum (x_i - \bar{x})(y_i - \bar{y}) / \sqrt{\sum (x_i - \bar{x})^2 \sum (y_i - \bar{y})^2}$). The results conveyed in \S \ref{sec:analysis} indicate the uncovered C$\equiv$N may be associated with larger PAHs, MAONs, and/or possibly endohedral or hetero fullerenes \citep[e.g.,][]{om16}. 

\begin{figure*}[!t]
\begin{center}
\includegraphics[width=0.85\linewidth]{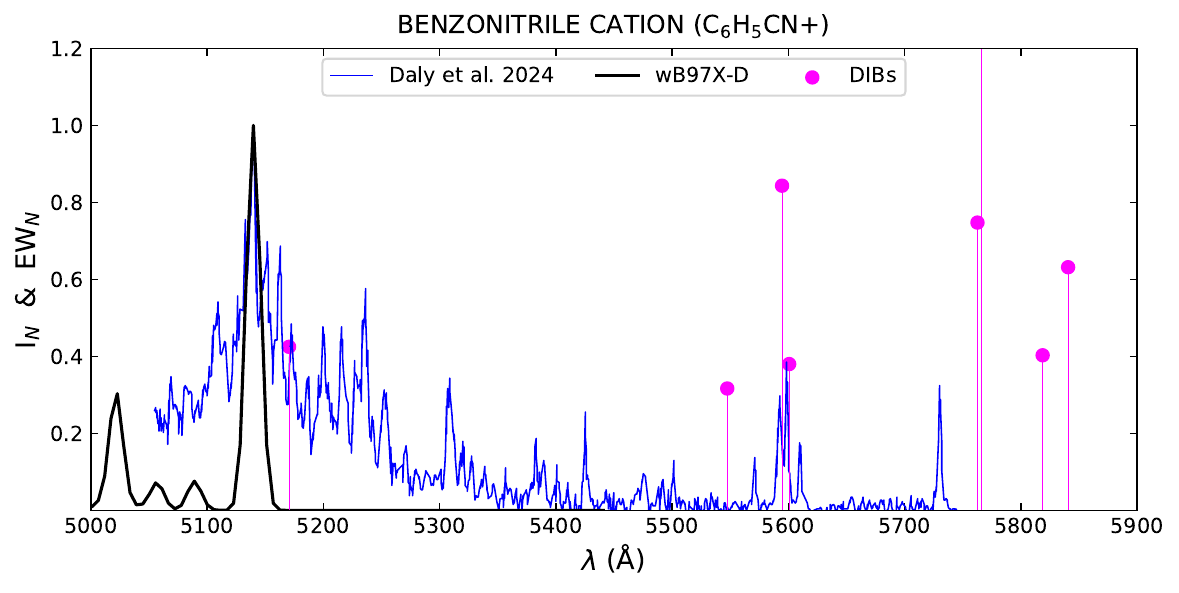}
\caption{Mismatches exist between observed DIBs (the expansive APO catalog), and $\omega$B97X-D/cc-pVTZ and experimental spectra. Experimental $\rm{C_6H_5CN^+}-{\rm{He}}$ data (blue line, merely broadly illustrative) were inferred from a diagram in \citet{da24}. APO DIBs shown exhibit $r\gtrsim0.8$ relative to 5600.72 {\AA} (EW pairs, $n\ge8$), and $|\Delta r| \lesssim 0.3$ between EW$-$E(B-V). Ultimately, solely two DIBs align with no less than ten experimental lines, which when paired with benzonitrile cation's IVR \citep{al96,mo13} and lack of recurrent fluorescence (\S \ref{sec-RF}), suggest in tandem with other evidence that the molecule is comparatively absent from the diffuse ISM.}
 \label{fig1}
\end{center}
\end{figure*}

In this study experimental, computational, and observational findings (e.g., the expansive APO DIB catalog) are mobilized to assess the comparative presence of $\rm{C_6H_5CN^+}$ in the diffuse ISM, consequently constraining molecular demographics in that specific environment by process of elimination. Experimental spectra are utilized that bypass matrices introducing sizable $\lambda$-shifts which complicate DIB matching \citep[e.g., neon matrix, tens of {\AA}ngstr{\"o}ms,][]{bp99}, hence the advantage of the \citet{da24} laboratory results exploited here.  By comparison DIBs can be on the order of {\AA}ngstr{\"o}ms \citep[e.g.,][the latter being the APO DIB catalog]{bon12,Fan2019}. Complementary synthetic spectra provide detailed characterizations of transitions (e.g., D0-D3, Fig.~\ref{fig1}), and the reader is referred to the following works for pertinent astrochemistry insights gleaned from quantum chemistry \citep{for21,zap23,esp24}. This initiative stemmed from a broader effort to delineate families of interrelated diffuse interstellar bands (DIBs), and scrutinize candidate molecular carriers \citep[e.g., \citealt{se25}, see also][]{bon12,fra21,esmith2022,ebe24,om24}.  However, false positives are common (e.g., molecules increasing in tandem), and a mitigation strategy is to employ parts of a challenging multidimensional approach ($r$ EW pairs, $|\Delta r|$ EW-$\rm{E(B-V)}$, potentially spectral morphology and FWHM, origin-band approach, optimal experimental spectra, etc., e.g., \citealt{gal08}).

\section{Analysis}
\label{sec:analysis}
The analysis aims to evaluate the detectability of $\mathrm{C_6H_5CN^+}$ in the diffuse ISM, which may lend further support to a top-down hypothesis (excluding ion-dipole reactions), or a bottom-up pathway where aromatic molecules might assemble in such conditions.

Interrelated DIBs possibly stemming from a common carrier were identified on the basis of equivalent widths (EWs) exceeding Pearson $r\gtrsim0.8$, relative to a DIB selected as the normalizing line ($\lambda_{E}\simeq5602.0$ {\AA}, Table 1 in \citealt{da24}, not He tagged).  That laboratory feature was chosen owing to its proximate wavelength to DIB 5600.72 {\AA}, and focused characterization by \citet{da24}. To mitigate false positives whereby highly correlated DIBs are tied to separate molecules whose abundances are commensurate, an absolute difference criterion was adopted between their EW correlations relative to optical reddening E(B-V) ($|\Delta r| \lesssim 0.3$). DIBs tied to the same molecular carrier should share a common correlation with reddening, and not necessarily a high correlation. The \citet[][]{Fan2019} APO DIB catalog was analyzed, and only EWs with uncertainties were assessed, in tandem with a minimum sightline sampling criterion ($n\ge8$). Unweighted correlations were evaluated.  The sightline to VI Cyg 12 was excluded since it features excess reddening beyond the field and may host a circumstellar shell \citep[e.g.,][]{Maryeva,xi24}.  For broader discussions regarding DIB EW and reddening correlations see \citet{fra21} and \citet{esmith2022}. 

Experimental $\rm{C_6H_5CN^+}$ spectra yield lines comparable in wavelength to DIBs at 5600.72 {\AA} ($\lambda_{E}\simeq5602.0$ {\AA}) and 5594.58 {\AA} ($\lambda_{E}\simeq5595.3$ {\AA}). Those peaks could shift depending on the temperature and other considerations \citep[][their Fig.~5 and Table 1]{da24}. The mean EWs for the DIBs are 3 and 6 m{\AA}, accordingly. EW ratios for 5594.58 and 5600.72 {\AA} are offset ($2.87\pm0.37$) from the near-unity that describes the experimental intensity ratios presented in Figs.~4 \& 5 in \citet[][]{da24}, and a high correlation ($r\simeq 0.9$) is juxtaposed upon a significantly offset $\Delta$EW-$\rm{E(B-V)}$ correlation, thereby favoring separate carriers increasing in commensurate fashion. The conclusion would benefit from independent data. The comparison is indirect between experimental intensity and observed EWs, and offsets are expected.  There is a general absence of matches elsewhere relative to observed EW and lab intensity ratios (Fig.~\ref{fig1}). The $\rm{C_6H_5CN^+}-{\rm{He}}$ data in Fig.~\ref{fig1} were inferred from a \citet{da24} diagram using PlotDigitizer, and provide a broadly qualitative proxy relative to $\rm{C_6H_5CN^+}$ \citep[comparisons in Table 1 of][]{da24}.  An APO DIB exists slightly blueward of 5600.72 {\AA} at 5599.76 {\AA} \citep[see also Table~3 in][]{da24}, with half the EW, and it is unclear if they represent one line \citep[e.g., DIBs 5779.59 and 5780.64 {\AA}, see][]{fra21}. A DIB at 5609.82 {\AA} is comparable to $\lambda_{E}\simeq5613.1$ {\AA}, yet the intensity to EW ratio differs from experimental spectra.  Nevertheless, in tandem with a low vibrational well to store excess photon energy \citep[e.g.,][]{al96}, Fig.~\ref{fig1} seemingly implies $\rm{C_6H_5CN^+}$ is unlikely to exist in detectable quantities in the diffuse ISM. That assertion is not tied to specific environments around emission-based celestial targets (UIEs).  

A generated spectrum is shown for the D0-D3 transition (Fig.~\ref{fig1}). Those simulated results were zero-pointed ($\lambda$) to the experimental D0-D3, and normalized to $\simeq 5140$ {\AA}. The experimental data do not sample the forbidden D0-D1 origin band, which is predicted near $\simeq 5803.9$ {\AA} \citep{xu06}.  The DIB observations are not complicated by telluric contamination or blinding as noted for the inconclusive C$_{70}^{+}$ case \citep[][see also the experimental data and interpretation of \citealt{cam16}]{ma25a}. \citet{da24} articulated that the absence of DIBs near $\simeq 5140$ {\AA} may arise from observational constraints (i.e., large FWHM).

Density functional theory was used to generate theoretical spectra, with the $\omega$B97X-D long-range corrected density functional and the Dunning correlation-consistent cc-pVTZ basis set. The ground state D0 (X$^2$B$_1$) of $\rm{C_6H_5CN^+}$ was optimized at this level of theory with a C2v symmetry point group, and frequency calculations were performed to ensure it constitutes a potential energy surface minimum.  The relevant electronic energy level scheme is shown as Fig.~3 in \citet{da24}. Time-dependent DFT (TD-DFT) at the same level of theory with 10 excited states was undertaken. From that analysis, and as \citet{da24} mentioned, D0-D1 and D0-D2 are both electronically forbidden.  An $f=0.1$ oscillator strength characterizes the D0-D3 transition ($\omega$B97X-D/cc-pVTZ). All calculations were performed via Gaussian 16 Rev.~C.02 \citep{g16}. The TD-DFT spectrum computed here supports the presence of symmetry-allowed and forbidden electronic transitions reported in the experimental work.

In sum, Fig.~\ref{fig1} juxtaposes DIBs from the APO catalog with an experimental $\rm{C_6H_5CN^+}-{\rm{He}}$ spectrum, and $\omega$B97X-D/cc-pVTZ results. Possibly two DIBs align with no less than ten experimental lines. Lowering the liberal $|\Delta r|$ EW-$\rm{E(B-V)}$ criterion results in even less overlap. 

\subsection{Improbable Recurrent Fluorescence}
\label{sec-RF}
\citet{mcg18} discovered that models were unable to explain the abundance of benzonitrile observed in TMC-1, a cold dense molecular cloud. The observed abundance was a factor of four larger than predictions, and hence pertinent formation pathways were missing. Subsequent work surmised that perhaps the shortfall could be addressed by benzonitrile cations in the diffuse ISM, and that fragmentation implied by the $\rm{C_6H_5CN^+}$ IVR could be bypassed via recurrent fluorescence \citep[see also the cyanonaphthalene research of][]{sto23}. \citet{da24} noted that hypothesis should be explored.

The lowest emissive electronic state should lie $<2.0$ eV above the ground state for plausible recurrent fluorescence to occur \citep{hau91,lac23}, and PAH cations associated with naphthalene, cyanonaphthalene, anthracene, and perylene exhibit $\lesssim 1.9$ eV gaps \citep{tok10,mar13,sai20,da23}. However, calculations here indicate that the lowest transition between two states for $\rm{C_6H_5CN^+}$ (D0-D3) is $2.54$ eV, which is analogous to the $2.58$ eV reported in Fig.~3 of \citet[][]{da24}.  Therefore, the $\rm{C_6H_5CN^+}$ transition is substantially higher than known recurrent fluorescence systems, and unlikely to be efficiently thermally populated. Moreover, that insight is paired with benzonitrile not adhering to a homogeneous trend in distribution maps of TMC-1 \citep{cer23}, which disfavors the cation as the key donor.  Benzonitrile in TMC-1 follows overdensities that likewise feature cyanopolyynes, and \citet{cer23} conclude that aromatics in TMC-1 are formed from smaller molecules in the densest regions of the cloud (i.e., not benzonitrile cations).  Furthermore, the smaller benzonitrile features far less vibrational modes than the aforementioned larger PAHs (inadequate energy diffusion via IVR), and will be struck by unceasing high-energy dissociating photons prior to or approximately concurrent with bottom-up encounters in the diffuse ISM.  Dissociative recombination is also a pertinent factor.

It has been suggested the lack of DIB matches to experimental $\rm{C_6H_5CN^+}$ arises because of the molecule's low oscillator strength and observational limitations, and that benzonitrile cations may be possibly abundant owing to recurrent fluorescence. A separate position is espoused here, whereby recurrent fluorescence is improbable in this instance, and based on the broader holistic evidentiary framework: the benzonitrile cation is relatively absent in the diffuse ISM.

\section{Conclusion}
\label{sec:conclusion}
The benzonitrile cation is comparatively absent in the diffuse ISM owing to a marginal vibrational diffusion mechanism for excess photon energy \citep{al96,mo13}, improbable recurrent fluorescence (\S \ref{sec-RF}), unceasing photon strikes, and general mismatches between experiment and observed DIBs (Fig.~\ref{fig1}). The findings in concert with the current lack of DIBs linked to comparably sized molecules reaffirm gaps inherent to bottom-up synthesis. Critically, low oscillator strength may inhibit the detection of molecules in the pathway sequence when solely relying on DIB matching, however, that limitation can be partially overcome by instituting the aforementioned holistic scaffolding, and by recognizing that presently solely C$_{60}^{+}$ is attributed to a few DIBs (see below). 

Future work entails examining whether the advocated falsification strategy is meritorious, and can be extended to intermediate-size aromatic nitriles. Essentially, the broader objective is detectability constraints spanning molecular size, environmental conditions and ionization state, recurrent fluorescence and IVR efficiencies. Concurrently, future efforts to characterize DIBs include ongoing research on fullerenes, and especially from the perspective of endohedral and hetero fullerenes \citep[e.g.,][and references therein]{om16}, and with respect to bolstering the C$_{60}^{+}$ case.  \citet{ma25a} unveiled that the existing controversy regarding the two most prominent DIBs linked to C$_{60}^{+}$ (9577 and 9632 {\AA}) likely stemmed from each team's data occupying a narrow dynamic range, and establishing a broad EW baseline confirmed the aforementioned lines exhibit a high correlation \citep[Fig.~1 in][]{ma25a}.  However, a match to C$_{60}^{+}$ cannot merely rest on two lines, and consequently \citet{maj35c} examined one of three weaker lines likewise debated in the literature, and produced amongst the first explicit Pearson correlation linked to a relevant sample size \citep[i.e., $9365-9577$ {\AA} are highly correlated, Fig.~2 in][]{maj35c}.  Yet, the \citet{maj35c} intercomparison of various published spectra revealed inhomogeneities leading them to agree with \citet{gal21} that evidence for the remaining two weaker lines was unconvincing ($9348-9428$ {\AA}), and \citet{maj35c} explicitly noted that additional research is needed on those two lines to mitigate false positives from overlapping PAHs (or potentially MAONs).  

Concerning C$_{70}^{+}$, a trio of \citet{cam16} experimental wavelengths match DIBs within an {\AA}ngstr{\"o}m \citep{ma25a}, however, numerous challenges underscored by \citet{ma25a} render that candidate carrier uncertain (see also the interpretation of \citealt{cam16}).  First, a suite of prominent lines could be obscured by telluric contamination \citep{cam16} and hinder the necessary full comparison, thus requiring (in)validation by space-based measurements \citep[e.g.,][]{co19}. Second, the low EWs complicate correlation analyses and indirect comparisons of observed EW ratios relative to experimental intensity attenuation ratios \citep[mismatched, see Table 4 and Fig.~2 in][]{ma25a}, especially since FWHM is challenging to determine in low-EW circumstances and certain lines may be incomplete or bifurcated \citep[see \S 2.2 in][]{ma25a}.  Work on all these diverse areas (\& more) is desirable.

\begin{acknowledgments}
This research relied on initiatives such as the APO DIB catalog, CDS, NASA ADS, arXiv, Gaussian 16, ChemCompute+GAMESS \citep{pw14}, and experimental spectra from the ISLA (Campbell) group. 
\end{acknowledgments}

\bibliography{article}{}
\bibliographystyle{aasjournalv7}

\end{document}